\documentclass[11pt,twoside]{article}
\usepackage{asp2010}
\usepackage[]{graphicx}

\resetcounters

\begin{document}

\title{The mystery of T\,Pyx; the 2011 explosion}
\author{Alessandro Ederoclite$^1$}
\affil{$^1$ Centro de Estudios de F\'isica del Cosmos de Arag\'on (CEFCA)\\
Plaza San Juan 1, Planta 2, 44001, Teruel, Spain
}

\begin{abstract}
T\,Pyx is a recurrent	nova which has	undergone eruptions on an almost regular basis
every 20 years until reaching a long lasting quiescence between 1967 and 2011. We observed
the long awaited 2011 explosion	 in the
optical and near infrared with	
intermediate spectral resolution. In
this paper we report	on	
the change in the spectral type of	
the nova (both during its rise and during its  fading), as  well as the observed	
 changes in the expansion velocities.
 We also present an interpretation of these changes and set them in the general	
framework of a new understanding of nova classification
\end{abstract}

\section{Introduction}

Recurrent novae (RNe) are a subclass of novae whose characteristics is to
have more than one observed explosion. This means that the recurrence time 
($\tau_\mathrm{R}$) is smaller than the $10^4$\,years which is expected 
for classical novae. Such short $\tau_\mathrm{R}$ is explained assuming a
massive white dwarf (approaching the Chandrasekhar mass) which would need
to accrete less mass in order to ignite thermonuclear reactions on its 
surface.

So far, about a dozen recurrent novae are known and they are divided in 
different classes: the T\,CrB/RS\,Oph class, the U\,Sco class and the T\,Pyx
class, based on the eruption properties (see \citealt{starrfield2008} and
\citealt{anupama2008}).

The ``T\,Pyx'' class is made of objects which look more like ``normal''
classical novae than like RNe. These objects have comparatively short 
periods (below the ``period gap'') and are classified as ``slow novae''.
Actually, only IM Nor and T\,Pyx itself fall in this class.

\begin{figure}
\includegraphics[width=\textwidth]{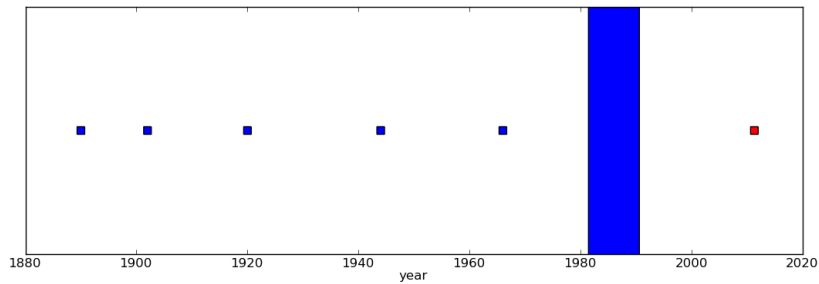}
\caption{
\label{fig:historic}
Historical eruptions of T\,Pyx. The shaded area is obtained by summing
to the epoch of the last eruption the average (plus or minus the standard
deviation) of the till-then recurrence times. }
\end{figure}

T\,Pyx was first observed in eruption in 1890 and, since then, it has 
had semi-regular eruptions (see Fig.\ref{fig:historic}).
The eruption predicted for the end of the 1980s has never been observed
thus giving rise to a series of speculations on the nature of this object
as well as on its next eruption
(see, e.g., \citealt{schaefer} and \citealt{selvelli} and references therein).
It was only recently \citep{uthas}, that the period of the binary was finally
measured: 1.8295\,h.
Making assumptions on the type of the secondary star, \citet{uthas} also
derived a mass for the white dwarf of $0.7 \pm 0.2$M$_\odot$.
which is much less than expected for a RN.

\section{The 2011 Eruption}

\subsection{The light curve}

\begin{figure}
\includegraphics[width=\textwidth]{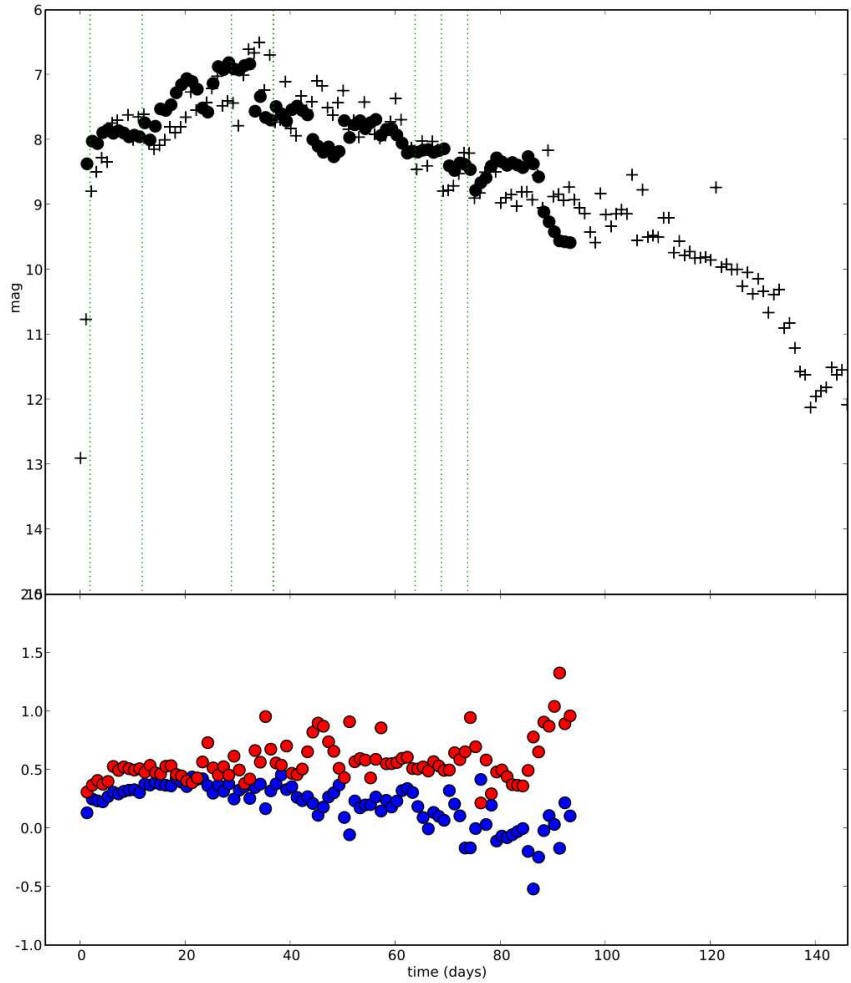}
\caption{\label{fig:lightcurve}AAVSO photometry of the 1966 and the 2011 eruptions of T\,Pyx. {\it 
Upper panel:} Crosses show the visual light curve of the 1966 eruption while
black dots show the B-band light curve of the 2011 eruption (time is in days
since discovery). {\it Lower panel:} Red points show the V-R color of the 2011
eruption while blue points show the B-V in the same eruption.}
\end{figure}

The long-awaited eruption of T\,Pyx occurred on Apr 14, 2011. It was reported 
by \citet{waagan}. Within 2\,days the nova brightened by 7\,mag. This phase
was followed by a ``plateau''. On day\,8, the light curve shows a last rise 
to maximum light ($V\sim6.5$\,mag) which occurred on day\,28.5.
The colors show some interesting evolution (see lower panel of 
Fig.\ref{fig:lightcurve}). The $B-V$ increases steadily from day\,0 to 
day\,20 and then start decreasing (although with some oscillation). The
$V-R$ increases only during the first 5 days and then stays (almost) 
constant throughout the evolution. Apparently, only after day\,80 the
$V-R$ clearly departs from this almost constant value. It is interesting to
observe a bump which is present both in the B-band plot and in the color
plots around day\,45. It may look like an episode of dust formation
but it has been observed \citep{evans+2012} that T\,Pyx has not formed dust 
(at least, not during this eruption).

It is interesting to make a comparison between the 1966 and the 2011 eruption.
Fig.\ref{fig:lightcurve} shows, in the upper panel, the visual band light
curve of the 1966 eruption and the B-band light curve of the 2011 eruption.
For clarity, both light curves have been re-sampled to a 1\,day. Although the
light curves are not identical (and taking into account the difference in 
bandpass), the light curves are remarkably similar, both showing the same
steep rise (although the 1966 light curve does not seem to show the same
pre-maximum plateau as the 2011 one) and the same slow decline.
The time between the 2011 eruption and its predecessor almost doubles
the time between the 1966 and its predecessor (see, again, Fig.\ref{fig:historic}). 
It is therefore quite 
remarkable that these two events show the same photometric characteristics,
unless one assumes that the mass transfer rate is reduced by a factor of 
almost two before and after the 1966 eruption.

\subsection{The spectrum}

\begin{figure}
\includegraphics[width=\textwidth]{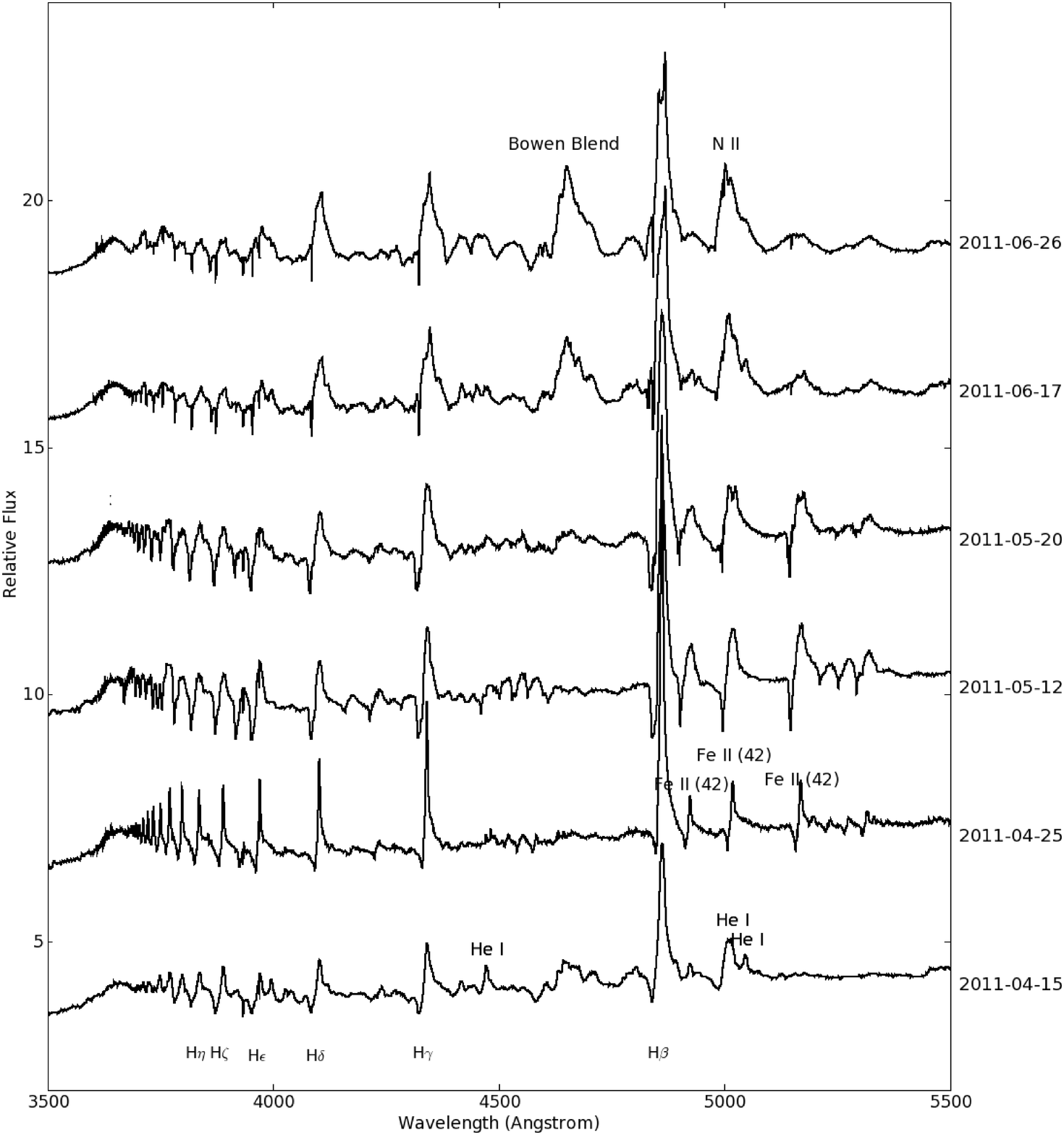}
\caption{\label{fig:spectra} Montage of the X-Shooter spectra obtained in the
framework of the present project. In order to be plotted in the same figure,
the spectra have been divided by their average value and then an arbitrary
offset has been added. Time goes from bottom to top. For clarity, only the UVB 
arms is shown. The rest of the spectra will be shown in a forthcoming paper.}
\end{figure}

The present spectra obtained with the X-Shooter spectrograph on the
Very Large Telescope between the day after the eruption and when the
star was too close to the Sun to be observed. X-Shooter is a multi-arm 
cross-dispersed 
echelle spectrograph. Two dichroics split the light in the three arms:
UVB, VIS and NIR, covering roughly 3000-5500\AA,5500-10000\AA, and 
10000-24000\AA, respectively. A montage of the UVB spectra is shown
in Fig.\ref{fig:spectra}.
The first spectrum that we present here was obtained on Apr 15, 2011, about 
26 hours after the report of the explosion.
The spectrum shows Balmer lines with P-Cyg profiles with expansion velocities
of about $1300 \pm 100$\,km~s$^{-1}$ and FWHM of $1130 \pm 200$\,km~s$^{-1}$.
For a comparison, on the day of discovery
\citet{izzo}, based on high resolution ($R = \frac{\lambda}{\delta \lambda}\sim57000$) spectra obtained with SARG on the
Telescopio Nazionale Galileo (TNG), reported also P-Cyg profiles with expansion velocities of
$\sim 1800$ km/s and FWHM $\sim 1200$ km/s.
This is consistent with what is expected for an ``Fe\,II'' nova
in the classification scheme by \citet{wms92}. Nevertheless, this
spectrum shows no Fe\,II lines but clear He/N lines. It therefore
falls in the ``He/N'' case as described by \citet{wms92}.

The second spectrum of our series was obtained 10\,days after the eruption.
The velocities (derived both through P-Cyg profiles and through
FWHM) are reduced by almost a factor of two. The appearance of the
spectrum is now completely different. The He and N lines are not 
detected and Fe\,II lines are clearly visible instead, thus 
making this spectrum a typical ``Fe\,II'' spectrum.

T\,Pyx reached its maximum brightness on day\,28.5 (see previous section).
The spectrum at this stage (on May\,12) shows Balmer lines, as well as 
Fe\,II lines and O\,I lines ($\lambda7773$\AA\ and $\lambda8446$\AA).
This is the typical spectrum of an ``Fe\,II'' nova at maximum light.
\citet{vlti}, based on near infrared interferometric data, has observed 
that, at this epoch, the expansion is bipolar. This does not show up
in our spectra where the P-Cyg profile clearly only shows a single absorption.
The expansion velocity, as measured from the FWHM, has started to increase
again (the minimum being reached at the previous epoch).

The spectra between day\,10 and day\,60 do not differ much from the
evolution of an Fe\,II nova. The P-Cyg profiles of the spectrum taken on
May 20 show multiple structures. This has already been observed in
other novae (e.g. V5114 Sgr, see \citealt{ederoclite}).
The spectrum taken on Jun 21, 2011
(day\,65) shows the emergence of the $[$N\,II$]$ line, thus 
suggesting the beginning of the nebular stage. Interestingly enough
no other forbidden line is observed. It is relevant to observe, instead,
that the Fe\,II lines are not visible (as expected in the evolution of
an Fe\,II nova evolving to the nebular stage) but 
several He\,I and N\,II lines are observable, hence, suggesting
that T\,Pyx has evolved back in an ``He/N state''. 
This transition ``backwards'' has already been reported for other novae 
(e.g. LMC\,1988\#2, see \citealt{wms92}), yet 
it is unusual
the simultaneous presence
of He\,I, N\,II, $[$N\,II$]$ while the Balmer lines are still produced in 
a relatively dense wind (as shown by the P-Cyg profiles).

\section{Summary and Conclusions}

The RN T\,Pyx was observed to begin its sixth known eruption on Apr 14, 2011. 
This reached maximum light on May 13, 2011 at $V\sim6.5$\,mag and a 
$t_2\sim20$\,days.
The photometric evolution looks qualitatively comparable with the one of the
1966 eruption, thus implying that similar physical processes must be at work.
Assuming that the mass accreted for the ignition is roughly the same (i.e. 
the white dwarf does not change significatively its mass during an explosion),
this implies that the mass accretion rate must be decreasing. A similar 
conclusion has been drawn by \citet{schaefer2005}.

The spectroscopic evolution is highly unusual (as already mentioned
in \citealt{wms2012}). The nova evolves from a 
``He/N'' phase to an ``Fe\,II'' phase and then back to the ``He/N'' phase.
We suggest the naming ``hyper-hybrid'' for this type of behavior and suggest
that it is due to a change in the optical depth in the ejecta. We speculate 
that the first spectral type change may be common in novae but it is 
normally missed because of the 
short time it takes them
 to rise to maximum light. The
second change may actually be due to a large physical extension of the 
ejecta in the radial direction. The furthest part of the ejecta
has already low enough density that can emit forbidden lines whereas
the inner part of the ejecta is still optically thick as witnessed by the
P-Cyg profiles.

It should be mentioned that the minimum of the evolution of the expansion 
velocity is reached when the nova makes its first spectral type ``transition''
(hence suggesting that this is due to an inwards movement of the 
photosphere). 
Finally, the mentioned ``bump'' in the light curve (and in the 
color curves) between days 40 and 50 is right before the epoch when the
``delayed mass ejection'' invoked by \citet{nelson2012} should occur and
it is right before the ``second spectral type change''.  

Throughout this conference, T\,Pyx has shown to be always the exception 
to the rule.
It is hard to tell if the evolution of T\,Pyx is unique or if its slow
evolution has allowed us to probe phases which are otherwise common to
all novae (or, at least, to a type of nova). 
High cadence observations (at high spectral resolution) would be required
in order to tell the difference between T\,Pyx and other novae. Most 
importantly, T\,Pyx has evidenced our lack of understanding of the very
early phase (the rise to maximum) where very unusual and unexpected changes
have been observed, thus (again) showing the importance of fast response
in the spectroscopic follow-up of this type of object.

\acknowledgements 
Based on observations made with ESO telescopes at the La Silla Paranal
Observatory under programme ID 287.D-5011.
CEFCA is funded by the Fondo de Inversiones de Teruel, supported by both the 
Government of Spain (50\%) and the regional Government of Arag\'on (50\%). 
This work was partly supported by the Spanish Plan Nacional de
Astrononom\'ia y Astrof\'isica under grant AYA2011-29517-C03-01.
I am grateful to Robert Williams and Elena Mason for the collaboration 
throughout the whole project. I am also grateful to Steve Shore for the 
enlightening discussions during the conference.

\bibliography{}

\end{document}